\documentclass{aa}
\pdfoutput=1
\usepackage[varg]{txfonts}
\usepackage{psfig,graphicx,natbib, gensymb}

\begin{document}

\title{The Rossiter-McLaughlin effect reloaded: \\Probing the 3D spin-orbit geometry, differential stellar rotation, and the spatially-resolved stellar spectrum of star-planet systems}
\titlerunning{The Rossiter-McLaughlin effect reloaded}
\authorrunning{Cegla et al.}

\author{H.~M. Cegla\inst{1,2}
	\and C. Lovis\inst{1}
	\and V. Bourrier\inst{1} 
	\and B. Beeck\inst{3}
	\and C.~A. Watson\inst{2}
	\and F. Pepe\inst{1}
} 

\offprints{H. M. Cegla, \email{h.cegla@qub.ac.uk}}

\institute{Observatoire de Gen\`{e}ve, Universit\'{e} de Gen\`{e}ve, 51 chemin des Maillettes, 1290 Versoix, Switzerland
  \and Astrophysics Research Centre, School of Mathematics \& Physics, Queen's University, University Road, Belfast BT7 1NN, UK
  \and Max-Planck-Institut f\"{u}r Sonnensystemforschung, Justus-von-Liebig-Weg 3, 37077 G\"{o}ttingen, Germany}

\date{Received X ? 2015 / Accepted X ? 201X}

\abstract {When a planet transits its host star, it blocks regions of the stellar surface from view; this causes a distortion of the spectral lines and a change in the line-of-sight (LOS) velocities, known as the Rossiter-McLaughlin (RM) effect. Since the LOS velocities depend, in part, on the stellar rotation, the RM waveform is sensitive to the star-planet alignment (which provides information on the system's dynamical history). We present a new RM modelling technique that directly measures the spatially-resolved stellar spectrum behind the planet. This is done by scaling the continuum flux of the (HARPS) spectra by the transit light curve, and then subtracting the in- from the out-of-transit spectra to isolate the starlight behind the planet. This technique does not assume any shape for the intrinsic local profiles. In it, we also allow for differential stellar rotation and centre-to-limb variations in the convective blueshift. We apply this technique to HD\,189733 and compare to 3D magnetohydrodynamic (MHD) simulations.  We reject rigid body rotation with high confidence ($>$99\% probability), which allows us to determine the occulted stellar latitudes and measure the stellar inclination. In turn, we determine both the sky-projected ($\lambda \approx -0.4\pm 0.2$$\degree$) and \textit{true 3D obliquity} ($\psi \approx 7^{+12}_{-4}$$\degree$). We also find good agreement with the MHD simulations, with no significant centre-to-limb variations detectable in the local profiles. Hence, this technique provides a new powerful tool that can probe stellar photospheres, differential rotation, determine 3D obliquities, and remove sky-projection biases in planet migration theories. This technique can be implemented with existing instrumentation, but will become even more powerful with the next generation of high-precision radial velocity spectrographs.
}



\maketitle

\section{Introduction}
\label{sec:intro} 
Amongst the numerous confirmed exoplanets, those that transit bright stars are the most highly prized. This is primarily because these targets can be observed both photometrically and spectroscopically; and the combination of the light curve and the radial velocity (RV) curve provides a more complete characterisation of the exoplanet system (e.g. obtaining both planet mass and radius). One additional, and extremely important, benefit of spectroscopically observing transiting planets is they allow us to measure the Rossiter-McLaughlin (RM) effect. The RM effect describes the RV anomaly that originates from the planet covering up regions of the stellar photosphere as it transits across the stellar disc. As the planet moves in front of the hemisphere that is rotating towards the observer, we observe a net redshift because the planet is obscuring a region of the star that is blueshifted due to the line-of-sight (LOS) component of the stellar rotation; and vice-versa as the planet transits the hemisphere rotating away from the observer. As such, the RM waveform is sensitive to the stellar rotation along the transit chord and therefore also sensitive to the alignment between the planet's orbital plane and the stellar spin-axis. 

Measuring this spin-orbit alignment is particularly important as it can feed into theories on planetary migration and evolution, which in turn underpins our understanding of planetary systems as a whole. For example, planets in aligned orbits may be indicative of a dynamically gentle planet-disc migration history, whereas misaligned planets may have experienced a more violent migration, such as planet-planet or star-planet scattering via the Kozai-Lidov mechanism. Such migration histories may in turn feed back into formation theories as some planets, such as hot Jupiters, are unlikely to have formed in their current observed locations. At present, there is evidence to suggest most cool stars ($T_{eff} < 6250$ K) host planets in aligned orbits, whereas hot stars may host planets with random alignments \citep[e.g.][and references therein]{winn10, albrecht13, brothwell14}. There is also tentative evidence that the star-planet misalignment may decrease with increasing system age \citep{triaud11}. Some authors argue that this dichotomy arises because stars with sufficient mass (and age) have a large enough convective envelope (and have had enough time) to realign planetary orbits through tidal dissipation \citep[see][for more discussions on potential star-planet dynamics]{brothwell14}. At present, additional obliquity measurements are needed to inform theories on the planet formation, evolution and migration for a variety of star-planet systems. 

Hence, correct modelling of the RM waveform is of critical importance to the exoplanet community. However, the current modelling of the RM waveform typically ignores any velocity source emanating from the stellar surface other than rigid body rotation. This is contradictory to our knowledge of stellar photospheres and may significantly impact our ability to accurately measure high precision RM observations. For example, we know that solar-type stars may exhibit differential rotation, that the intrinsic line profiles (and also observed cross-correlation functions [CCFs]) are asymmetric due to stellar surface granulation, and that these line profiles/CCFs experience velocity shifts due to both stellar oscillations and granulation. 

To date, a number of analytical and numerical RM models have been constructed. \cite{hirano10, hirano11} and \cite{boue13} constructed analytical expressions for the RM waveform for non-stabilised and stabilised spectrographs, respectively. However, in each case they neglected to account for differential rotation\footnote{Though \citealt{hirano11} did demonstrate that this effect can be significant, at least for fast rotating stars.}, and assumed constant, symmetric intrinsic line profiles/CCFs. Additionally, \cite{oshagh13} and \cite{shporer11} both constructed model stars in attempts to numerically model the RM effect. Again, these authors assumed a rigidly rotating stellar surface with the stellar photospheric lines approximated by constant Gaussian functions. However, \cite{shporer11} did account for the local convective blueshift to first order by treating the net convective velocity as a constant that varies across the stellar disc due to projected area (but neglected dependences on the centre-to-limb position); these authors were able to show this effect impacted the RV anomaly on the m~s$^{-1}$ level. There have also been attempts to directly analyse the distortions in the stellar spectra/CCFs caused by the planet occulting the stellar surface (rather than the disc-integrated RVs used in the previous models). \cite{cameron10} and \cite{albrecht13} have pioneered these techniques for star-planet studies with stabilised and non-stabilised spectrographs, respectively. In practice, both authors assumed the stellar photosphere produces constant, symmetric line profiles/CCFs and rotates rigidly; though, \cite{albrecht13} did attempt to account for the convection effects following the \cite{shporer11} approximation. Furthermore, the residuals for current RM models sometimes have a strange wave-like form that may indicate a more detailed modelling is warranted \citep[see e.g.][]{triaud09, brown15}

More recently, \cite{cegla15b} have shown that ignoring the centre-to-limb convective variations (in net blueshift and intrinsic profile asymmetry) across the stellar disc of a rigidly rotating Sun-like star may result in residuals (between observed and fitted data) with amplitudes of 10s of cm~s$^{-1}$ to $\sim$10~m~s$^{-1}$ for stars with $v_{eq} \sin i$ of 1-10~km~s$^{-1}$ (in the case of an aligned system with a 4 d hot Jupiter and an impact factor of 0). They also reported that neglecting these effects may cause observers to underestimate their errors on the projected obliquity by 10-20$\degree$, and that incorrectly modelling the intrinsic profile asymmetry for a moderately rapidly rotating star (with a $v_{eq} \sin i_{\star}$ = 6~km~s$^{-1}$) may cause systematic errors on the measured obliquities that are incorrect by $\sim$20-30$\degree$. Consequently, for systems where the stellar photospheric lines cannot be approximated by a Gaussian and/or the centre-to-limb convective variations cannot be ignored, the typical RM modelling may systematically bias our interpretations of the RM waveform. 

Additionally, ignoring significant (solar-like) differential stellar rotation could lead to an underestimation of the $v_{eq} \sin i_{\star}$ reported through RM modelling. This could potentially contribute to the known discrepancy between the $v_{eq} \sin i_{\star}$ reported by spectral line broadening and that reported by RM modelling \citep[see e.g.][]{triaud15}. For systems with a star-planet misalignment, neglecting differential rotation could also lead to biases in the derived obliquities; this is because a latitudinal dependency on rotation could be misinterpreted as a difference in star-planet alignment. Furthermore, if rigid-body rotation can be excluded at high confidence and the system is even slightly  misaligned, then accounting for differential rotation allows us to lift the degeneracy between the equatorial velocity and the stellar inclination. This is because the transit is then sensitive to the stellar latitudes occulted by the planet \citep[as noted by][]{gaudi07}, which allows us to directly measure the stellar inclination. If we can lift the $v_{eq} \sin i_{\star}$ degeneracy, we can measure the \textit{true 3D spin-orbit geometry} of the star-planet system. This in turn can help remove biases introduced by studying the \textit{sky-projected} obliquities in planet migration theories. 

For these reasons, in this paper, we seek a RM modelling technique that allows for differential stellar rotation, does not assume any particular function for the shape of the intrinsic stellar photospheric lines, and allows the centre-to-limb convective variation to contribute to the observed RV anomaly. Throughout this paper, we apply a new RM modelling technique to high precision HARPS observations of the transit of HD\,189733\,b. We also compare these empirical results to radiative 3D magnetohydrodynamic (MHD) simulations of a K dwarf (in a manner similar to \citealt{dravins14}); such a comparison allows us to test the realism of 3D MHD simulations for non-solar main sequence stars for the first time and can provide insight into the underlying physics of the observed stellar photosphere. 

In Section~\ref{sec:setup}, we describe how transiting planets can be used to probe (and effectively resolve) the stellar photosphere by isolating the light behind the planet along the transit chord. Here, we present the observed and simulated data, as well as an overview of the `planet-as-a-probe' technique. In Section~\ref{sec:results}, we provide measurements on the differential stellar rotational velocity, 3D spin-orbit geometry, and the net convective velocity shifts for  HD\,189733; we also present our findings on the observed CCF and simulated line profile changes across the stellar disc. We conclude and discuss the significance of these findings in Section~\ref{sec:conc}. 

\section{Planet-as-a-Probe: \\ Resolving the Stellar Surface}
\label{sec:setup}
When a planet transits its host star, the stellar photosphere behind the planet is blocked from the line-of-sight (LOS). We can isolate the starlight from these occulted regions by subtracting in-transit spectroscopic observations from those taken out-of-transit. This technique is currently used in line profile tomography, pioneered for exoplanet studies by \cite{cameron10}. For a stabilised spectrograph, this analysis is performed on the observed CCF. The out-of-transit CCF (CCF$_{out}$) is modelled by a rotationally-broadened, limb-darkened Gaussian convolved with the spectrograph's instrumental profile. The in-transit CCF (CCF$_{in}$) is then modelled by the addition of a travelling Gaussian `bump' (due to the planet presence) to the out-of-transit profile. The spectral position of this `bump' depends on the planet position on the stellar disc (as seen by the observer), and its amplitude is proportional to the fraction of starlight obscured by the planet; while the Gaussian itself represents the average stellar photospheric CCF behind the planet. Hence, this technique allows one to model the missing starlight using the planet-as-a-probe and provides a way to resolve the stellar surface along the transit chord (see \cite{cameron10} for more details). 

In this paper, we employ a similar technique to isolate the starlight behind the planet (described in detail in Section~\ref{subsec:tech}), but we do not assume a particular function for the local CCF in order to model its impact on the disc-integrated CCF. Instead, we analyse directly the local CCF occulted by the planet. This means we do not have to assume any particular shape for the local CCF. We also go further and allow for both differential rotation and centre-to-limb net convective variations.

\begin{table}[b!]
\caption[]{Fixed parameters for HD\,189733}
\begin{center}
\begin{tabular}{c|c|c}
    \hline
    \hline
      Par. & Value & Reference \\
    \hline  
	$T_{0}$ & 2454279.436714 $\pm$ 0.000015 d & \citealt{agol10} \\
	$P$ & 2.21857567 $\pm$ 0.00000015 d & \citealt{agol10}\\
	$i_p$ & 85.710 $\pm$ 0.024$\degree$ & \citealt{agol10} \\
	$R_{\star}$ & 0.805 $\pm$ 0.016 R$_{\odot}$& \citealt{boyajian15}\\
	$R_p$ & 0.15667 $\pm$ 0.00012 R$_{\star}$ & \citealt{sing11}\\
	$a/R_{\star}$ & 8.863 $\pm$ 0.020 & \citealt{agol10} \\
	$T_{dur}$ & 1.827 hr & \citealt{torres08} \\
	$K$ & 200.56 $\pm$ 0.88 m s$^{-1}$ & \citealt{boisse09} \\
	$e$ & 0 & \\
	$\omega$ & 90$\degree$ & \\
	$u_1$ & 0.816 $\pm$ 0.019 & \citealt{sing11} \\
	$u_2$ & 0 & \citealt{sing11} \\
	$u_1$ & 0.548 & simulation\tablefootmark{a}\\
	$u_2$ & 0.213 & simulation\tablefootmark{a} \\
	T$_{eff}$ & 4875 $\pm$ 43 K & \citealt{boyajian15}\\
	log $g$ & 4.56 $\pm$ 0.03 & \citealt{boyajian15}\\
    \hline
  \end{tabular}
\end{center}
\tablefoot{\tablefoottext{a}{The radiative MHD simulations naturally include limb darkening, and as a result these values were obtained by fitting the limb darkening function to the simulated line profiles.}}
\label{tab:fix}
\end{table}

\subsection{Observational Data}
\label{sub:obs}
In order to use the planet-as-a-probe to resolve the stellar surface, we required a bright target, with high signal to noise, observed on a highly stabilised spectrograph. The obvious first choice was HD\,189733 due to the availability of archival observations from the HARPS (High-Accuracy Radial-velocity Planet Searcher) echelle spectrograph on the ESO 3.6 m telescope in La Silla, Chile. In total, there are four nights of data: two from 2006 (July 29/30 and September 7/8) and two from 2007 (July 19/20 and August 28/29). This data set, and subsets of it, have been studied a number of times and further details on the observations can be found in \citet{triaud09}, \citet{cameron10}, and \citet{wyttenbach15}. The exposure times range from 300-900 s (highest in-transit exposure is 600 s) and the signal-to-noise ratio extracted per pixel ranges from 100-170 in the continuum near 590 nm \citep{wyttenbach15}. Only half a transit was observed on July 29, 2006 due to poor weather in the second half of the night. In total, there are 111 CCFs, with roughly half the CCFs observed in-transit. Throughout our analysis, we operate on the CCF output by the HARPS pipeline, created by an order-by-order cross-correlation of the stellar spectrum with a standard mask function; the mask function was weighted by the depth of the lines, wherein the lines were derived from those observable in the spectrum of Arcturus (excluding telluric regions). Since HD\,189733 has been observed extensively, beyond RM measurements, we fix many system parameters to their literature values; these can be found in Table~\ref{tab:fix}.

\subsection{Technique Overview}
\label{subsec:tech}
To begin, we removed the Doppler-reflex motion induced by the presence of the planetary companion; this was done assuming a circular orbit with a semi-amplitude provided by \cite{boisse09}, who modelled the full RV curve. We then created four master out-of-transit CCF$_{out}$ by co-adding all the out-of-transit CCFs together for each given night. Separating the CCF$_{out}$ for each run allowed us to directly account for nightly offsets in the instrumental, atmospheric, and astrophysical noise. In-line with this, the RV from each master CCF$_{out}$ was removed from all the individual CCFs, on a night-by-night basis.

\begin{center}
\begin{figure}[t!]
\centering
\includegraphics[trim=0.49cm .25cm 0.cm 1cm, clip, scale=0.44]{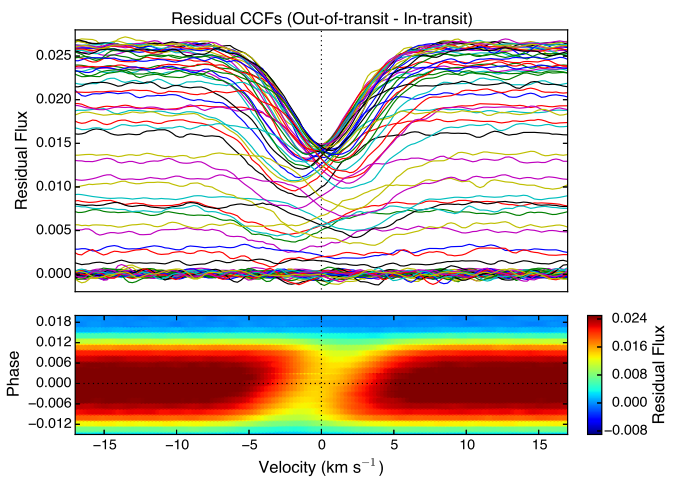}
\caption[]{Top: Residual CCF profiles; colours chosen for viewing ease. Bottom: Residual map of (a subset of) the time series CCFs, colour-coded by residual flux. Dotted lines at 0 phase and 0 km s$^{-1}$ are shown to guide the eye. The travelling `bump' in the CCF$_{in}$ profiles is evident as a bright streak, as the planet traverses the stellar disc. The transit centre occurs at $\sim$0 km s$^{-1}$ in both plots because the orbital motion and systemic RVs were removed.} 
\label{fig:resid}
\end{figure}
\end{center}
\vspace{-30pt}

Since the HARPS observations are not calibrated photometrically, the continuum flux of the individual CCFs is on an arbitrary scale. To compare in- and out-of-transit CCFs, we first scaled the continuum flux according to a Mandel \& Agol transit light curve with a quadratic limb darkening law (using the input values given in Table~\ref{tab:fix} -- note the limb darkening coefficients for the observations were chosen following the white light HST STIS light curve fit from \cite{sing11}; however, we did test the impact of assuming the limb darkening coefficients for the R and B bands, but found no significant difference in the final results\footnote{The best-fit parameters agreed within 1-2$\sigma$ with those in Section~\ref{sec:results}.}). This scaling allowed us to directly subtract all the CCF$_{in}$ from their respective master CCF$_{out}$ to isolate the starlight behind the planet. Directly subtracting the in-transit from the out-of-transit means we do not have to assume any shape for the residual CCFs (i.e. the local regions of stellar photosphere occulted by the planet). 

\begin{center}
\begin{figure*}[t!]
\centering
\includegraphics[trim=0.5cm 0.44cm 0.25cm 0.4cm, clip, scale=0.945]{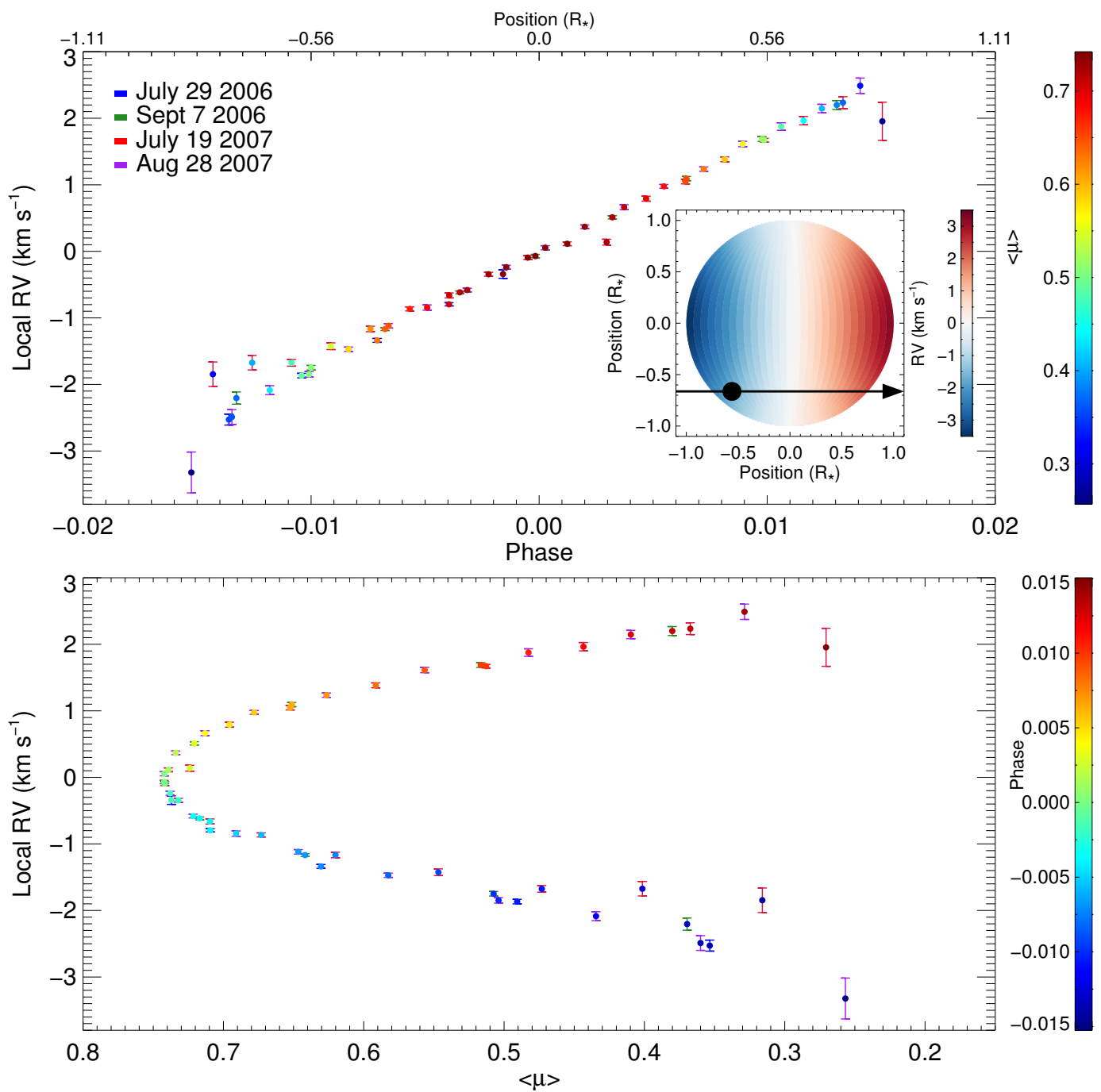}
\caption[]{Top: Net velocitiy shifts of the in-transit residual CCF profiles (i.e. CCFs of the regions occulated by the planet) as a function phase (bottom axis) and stellar disc position defined in units of stellar radii (top axis), with an inset illustrating the planet positions across the star; the data are colour-coded by the disc position in units of the brightness-weighted $\left < \mu \right >$ (where $\mu = \cos \theta$) behind the planet, while the colour of the error bar indicates the observation date. Bottom: Same velocities, but plotted against $\left < \mu \right >$ and colour-coded by phase.
}
\label{fig:netvel}
\end{figure*}
\end{center}
\vspace{-30pt}

The residual CCFs are shown in Figure~\ref{fig:resid} (different colours used for viewing ease); as a reminder these are set in the stellar rest frame because the nightly systemic velocities of the master CCF$_{out}$ have been removed. The stellar rotational velocity is clearly evident in the residual CCFs. Residual CCFs near the limb have lower continuum flux due to the assumed limb darkening (profiles at ingress/egress have an additional drop in flux since the planet is only partially on the stellar disc) and higher velocity shifts due to stellar rotation. The velocity shifts of the profiles nearest the limb are $\sim$ 3 km s$^{-1}$; this is in agreement with the $v_{eq} \sin i_{\star}$ reported in the literature, which ranges from $\sim$ 2.9 - 3.5 km s$^{-1}$ (see Table~\ref{tab:vsini}). 

Throughout the paper, RVs were determined from the mean of a Gaussian fit to the CCFs, as this one of the most standard RV determination techniques for data taken with a stabilised spectrograph. We performed the fit on one in four points, due to the over-sampling of the CCF output by the HARPS pipeline, using a Levenberg-Marquardt least-squares minimisation \citep[][and references therein]{markwardt09}. The flux errors assigned to the CCFs were derived from the standard deviation in the continuum flux; the error from the Gaussian fit corresponds to the one sigma statistical errors calculated from the square root of the diagonal elements of the covariance matrix (output from the minimisation). The total velocity shifts of the CCFs are directly measured and therefore include contributions from both the stellar rotation and any net convective velocities (as well as any other velocity sources that may originate on the stellar surface).

The net velocity shifts of the in-transit residual CCFs are shown in Figure~\ref{fig:netvel}; in the top plot they are plotted against phase and in the bottom plot they are plotted against the stellar disc position, defined as the brightness-weighted $\left < \mu \right > $ behind the planet. This was computed numerically as 
\begin{equation}
\label{eqn:mu}
{ \left < \mu \right >} = \frac{\sum I \ \mu}{\sum I},
\end{equation}
where  $\mu = \cos \theta$ ($\theta$ is the centre-to-limb angle) and $I$ is the intensity determined from the aforementioned quadratic limb darkening. To compute $\left < \mu \right >$ numerically, we constructed a stellar grid that is transited by the planet (see Section~\ref{subsubsec:stellrot} for more details on this grid). The summation over the $\mu$ behind the planet is performed over a square grid, defined with an origin at the planet centre, in 51 equal steps in the vertical and horizontal direction (note we varied the number of steps used to compute the average and found no major difference when using more pairs). Contributions from steps that do not lie beneath the planet and/or on the stellar disc were excluded. We removed CCFs with $\left < \mu \right > <$ 0.25 from our analysis as profiles close to the limb were very noisy. The data points are colour-coded by phase, while the colour of the error bars indicates the observation date. In the bottom of Figure~\ref{fig:netvel}, these velocities are plotted against phase (colour-coded by $\left < \mu \right > $, with same error bars) and shown alongside a schematic of the transit to further illustrate the location and velocities of the residual CCFs.  

As can be seen in Figure~\ref{fig:netvel}, many points in the July 19 data are shifted relative to the other nights at similar stellar disc positions (e.g. the data on the first half of the transit and a datum near disc centre do not follow the RV trends seen in the rest of the data). The most likely culprit for the offset July 19 data is that during this night the planet transits regions of different magnetic field strength. For example, an increased magnetic field strength will inhibit the convective flows and therefore may affect the net convective blueshift. To explore the potential level of magnetic activity for each night, we examined the observed log~$R_{HK}$ for each in-transit data point. Unfortunately, the precision of the log~$R_{HK}$ was not sufficient to draw succinct conclusions; however, the log~$R_{HK}$ does potentially suggest that on July 19 the star may be less magnetically active (with a log~$R_{HK} \approx -4.53$, compared to the other nights with log~$R_{HK} \approx$ -4.52 to -4.48). Additionally, comparing to the trends in MHD simulations suggests that some of the shifted July 19 data may actually be due to the planet occulting regions of less magnetic activity (see Section~\ref{subsec:CB} for more details on the relationship between magnetic field and centre-to-limb convective variations). Nonetheless, regardless of the origin of this trend, we excluded all of the July 19 data to avoid biasing the final results. 

We add a cautionary note that stellar activity may have biased previous RM measurements, especially if the planet occults regions of vastly different activity (whether that be of increased or decreased magnetic field). However, our technique may have the advantage of making variations in the stellar activity more apparent since we analyse directly the local phoptophseric CCFs. Hence, this technique may present a unique opportunity to study the properties of active regions and localised surface flows on other stars.

\subsubsection{Modelling the Doppler-shifts of the Residual CCFs}
\label{subsubsec:stellrot}
In order to model the RM waveform, we account for the Doppler-shifts of the residual CCF profiles due to the stellar rotation behind the planet and allow for additional shifts from centre-to-limb convective variations. An assumption of rigid body stellar rotation could systematically bias the rotation contribution across the stellar disc if significant differential rotation is present. For HD\,189733, \cite{fares10} report a latitudinal angular velocity shear of d$\Omega$ = 0.146 $\pm$ 0.049 rad d$^{-1}$. Combining this  with their reported equatorial rotation, $\Omega_{eq}$ = 0.526 $\pm$ 0.007 rad d$^{-1}$, yields a relative differential rotation rate of $\alpha$ = 0.278 $\pm$ 0.093 (where $\alpha$ = d$\Omega/\Omega_{eq}$). A differential rotation of this magnitude would differ from rigid body rotation on the 100s of m s$^{-1}$ level. Additionally, \cite{beeck13} reported centre-to-limb convective velocity shifts for Fe\,\textsc{i} line profile cores from 3D MHD simulations of a K0V dwarf on the order of $\le$ 100 m s$^{-1}$. Hence, allowing for differential rotation may be the only way for us to determine the convective contribution to the measured RVs of the residual CCFs; moreover, it may also be the only way for us to directly determine the stellar latitudes transited and hence disentangle the true 3D spin-orbit geometry from projection effects. 

\begin{center}
\begin{figure}[t!]
\centering
\includegraphics[trim=7.7cm .1cm 5.cm 4.cm, clip, scale=0.35]{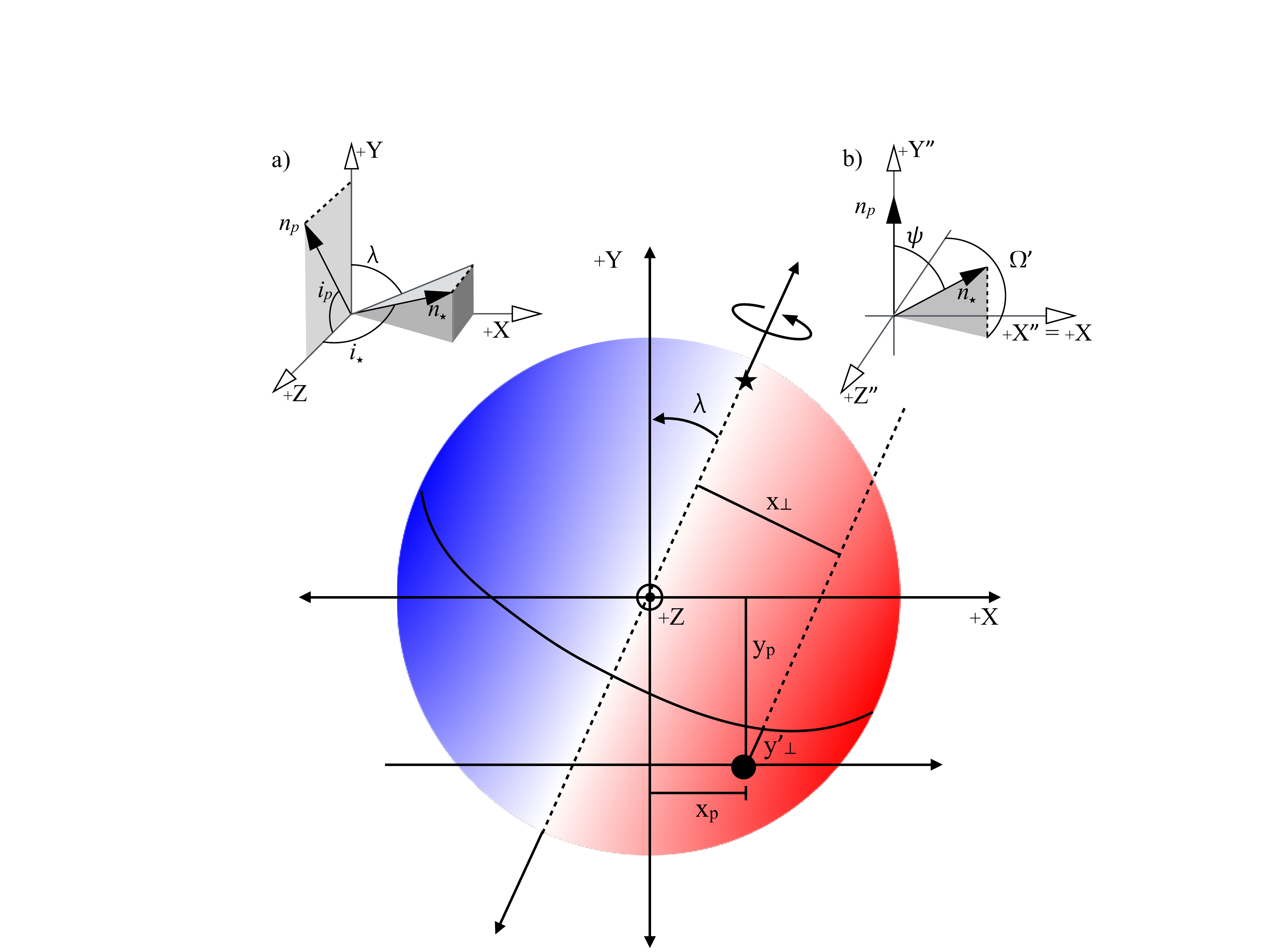}
\caption[]{Schematics of the coordinate system. Main: 2D projection of the system, illustrating the observed planet centre ($x_p, y_p$), the orthogonal distances from the stellar spin-axis and equator ($x_{\perp}, y_{\perp}'$ -- respectively), and the projected obliquity ($\lambda$); the stellar pole is indicated by a star. Insets a) and b) are adapted from \cite{fabrycky09}, to illustrate the 3D observer-oriented and orbit-oriented reference frames, respectively. Inset a): illustrates the projected obliquity in relation to the orbital inclination ($i_p$), stellar inclination ($i_{\star}$), and the normals to the orbital plane ($n_p$) and stellar rotation ($n_{\star}$). Inset b): (XYZ)$''$ obtained after a rotation of the observer-oriented frame about the x-axis by $\pi/2 - i_p$, illustrates the 3D obliquity ($\psi$) and the azimuthal angle ($\Omega'$ -- not to be confused with the angular rotation denoted earlier as $\Omega$).} 
\label{fig:refframe}
\end{figure}
\end{center}
\vspace{-30pt}

To model the stellar rotation contribution to the residual CCF velocities, we computed the brightness-weighted average differential rotation behind the planet for each observational epoch ($v_{ \left < stel \right >}$). The centre of the planet at any given orbital phase, $\phi$, can be described as

\begin{equation}
\label{eqn:x}
x_p = \frac{a}{R_{\star}} \sin (2\pi \phi),
\end{equation}

\begin{equation}
\label{eqn:y}
y_p = -\frac{a}{R_{\star}} \cos (2\pi \phi) \cos(i_p),
\end{equation}
for a circular orbit, where $a$ is the orbital semi-major axis, $R_{\star}$ is the stellar radius, and $i_p$ is the orbital inclination (with the y-axis parallel to the projected direction of the orbital axis, and $+$z-axis pointing towards the observer -- see Figure~\ref{fig:refframe}). For differential rotation, we are interested in the orthogonal distances from the stellar spin-axis and equator. The orthogonal distance from the spin-axis ($x_{\perp}$) can be determined by rotating our coordinate system in the plane of the sky by the projected obliquity, $\lambda$, yielding

\begin{equation}
\label{eqn:xperp}
x_{\perp} = x_p \cos (\lambda) - y_p \sin (\lambda),
\end{equation}

\begin{equation}
\label{eqn:yperp}
y_{\perp} = x_p \sin (\lambda) + y_p \cos (\lambda), 
\end{equation}
for the centre of the planet; the orthogonal distances for any given point can be calculated by replacing $x_p$ and $y_p$ with a given $x$ and $y$. However, this rotation does not guarantee alignment between the coordinate reference system equator and stellar spin equator, unless $i_{\star} = 90\degree$. This is important as we need the orthogonal distance from the stellar equator in order to calculate the differential rotation. To obtain this ($y_{\perp}'$), we then further rotated our coordinate system about the $x_{\perp}$ axis (in the $z_{\perp}y_{\perp}$ plane) by an angle $\beta = \pi/2 - i_{\star}$:
\begin{equation}
\label{eqn:zperpp}
z_{\perp}' = z_{\perp} \cos (\beta) - y_{\perp} \sin (\beta),
\end{equation}

\begin{equation}
\label{eqn:yperpp}
y_{\perp}' = z_{\perp} \sin (\beta) + y_{\perp} \cos (\beta), 
\end{equation}
where $z_{\perp} = \sqrt{1 - x_{\perp}^2 - y_{\perp}^2}$ (note $x_{\perp}' = x_{\perp}$ since we rotate about the $x_{\perp}$ axis), since we defined our coordinate system in units of $R_{\star}$. 

Then by assuming a differential rotation law derived from the Sun ($\Omega = \Omega_{eq}\ (1 -  \alpha \sin^2 \theta)$), the stellar rotational velocity for a given position is defined as 

\begin{equation}
\label{eqn:vstel}
v_{stel} = x_{\perp} v_{eq} \sin i_{\star} (1 -  \alpha~y_{\perp}'^2),
\end{equation}
since $y_{\perp}'$ = $\sin (\theta_{lat})$, where $\theta_{lat}$ is the latitude relative to the stellar equator and $\alpha$ is the aforementioned differential rotation rate. From the planet centres, we were also able to compute $x_{\perp}$ and $y_{\perp}'$ for any number of $x,y$ positions on the stellar disc and behind the planet. We then used these in numerically determining the brightness-weighted average value stellar rotational velocity occulted by the planet:

\begin{equation}
v_{ \left < stel \right >} = \frac{\sum I \ v_{stel}}{\sum I}.
\end{equation}
Similar to Equation~\ref{eqn:mu}, the summation is performed over 51 equal steps in $x$ and $y$, centred on $x_p$,~$y_p$ (and contributions from steps that do not lie beneath the planet and/or on the stellar disc were set to 0). Note we also tested including the effect of finite exposure time in $v_{ \left < stel \right >}$ by averaging together the $v_{ \left < stel \right >}$ at the start, middle, and end of the exposure. However, including this effect resulted in differences which were significantly less than the errors on the RVs of the residual profiles. Hence, we proceeded to use only the average values calculated at the time of mid-exposure. Additionally, we also tested the rigid body stellar rotation assumption by fixing $i_{\star} = 90\degree$ and $\alpha = 0$ in Equation~\ref{eqn:vstel}. 

Note that the total velocity that we measure for the CCF behind the planet also includes velocity shifts due to stellar surface magneto-convection. For this paper, we exclude the temporal variability induced by granulation as the precision of the data does not allows us to constrain this small-scale phenomenon. Instead, we include only the larger centre-to-limb variability, due to the corrugated nature of granulation. Hence, the total measured velocity is \begin{equation}
\label{eqn:vtot}
v_{tot} = v_{ \left < stel \right >} + v_{conv},
\end{equation}
where $v_{conv}$ is the net convective velocity shifts. We do not know an exact formulation for the convective contribution, $v_{conv}$. To circumvent this, we approximate the convective contribution using a polynomial; we tried a variety of polynomials, from zeroth to second order (note we did attempt a third order polynomial, but found the data was insufficient to constrain such a high order polynomial). Due to the corrugated nature of granulation on the spherical host star, $v_{conv}$ will be radially symmetric about the disc centre. Hence the centre-to-limb convective velocity contribution is defined as

\begin{equation}
\label{eqn:vconv}
v_{conv} = \sum_{i=0}^{i=n} c_i \left < \mu \right >^i,
\end{equation}
where $n$ is the polynomial order. However, because we previously removed the RVs from the master out-of-transit CCFs we effectively removed, from our in-transit data points, the brightness-weighted net convective velocity of the whole stellar disc. Hence, when fitting the $v_{conv}$ polynomial to our in-transit data, the coefficients of the polynomial must be such that the brightness-weighted net CB integrated over the stellar disc is equal to zero, i.e. 

\begin{equation}
\label{eqn:frac0}
\frac{\int_0^\pi 2\int_0^{\pi/2}  I (\theta) \ v_{conv}(\theta) \ R_{\star}^2 \sin(\theta) \ d\theta d\phi}{\int_0^\pi \int_0^{\pi/2} I(\theta) \ R_{\star}^2 \ \sin(\theta) \ d\theta d\phi} = 0,
\end{equation}
where the surface element $dS_{R_{\star}} = R_{\star}^2 \sin(\theta) d\theta d\phi$. The integration is performed from $\phi$ from 0 to $\pi$ because we are only interested in the half of the sphere facing us, and the integration over $\theta$ can be written as twice the integral from 0 to $\pi/2$ because both halves of the stellar disc are considered equal (since the centre-to-limb variation is radially symmetric). By rewriting Equation~\ref{eqn:frac0} in terms of $\mu$, inserting  Equation~\ref{eqn:vconv}, and solving for the constant offset ($c_0$) in $v_{conv}$, we find: 
\begin{equation}
\label{eqn:co}
c_0 = - \frac{\sum_{i=1}^{i=n} c_i  \int_0^1 I(\mu)\ \mu^{i+1} \ d\mu}{\int_o^1 I(\mu)\ \mu \ d\mu}. 
\end{equation}
Hence, any $v_{conv}$ polynomial determined herein must satisfy Equation~\ref{eqn:co} (since we previously removed the nightly net out-of-transit convective velocity shift).  

It is very probable that any transiting planet suitable for the analysis described herein will also have high precision light curves from which $a/R_{\star}$, $i_p$, and $R_p/R_{\star}$ can be determined more accurately and precisely than from the Rossiter-McLaughlin measurements alone. This is the case for HD\,189733 (see Table~{\ref{tab:fix}), which we utilised in order to compute $x_p$ and $y_p$ for each in-transit epoch. To compute the brightness-weighted average rotational velocity behind the planet also requires $v_{eq}$, $ i_{\star}$, $\lambda,$ and $\alpha$. We fit for these quantities using a Metropolis-Hasting Markov chain Monte Carlo (MCMC) algorithm. Allowing for a non-zero $v_{conv}$ contribution also means our MCMC must fit for each of the coefficients given in Equation~\ref{eqn:vconv}. We do not presume any a priori knowledge of $v_{conv}$, except that it must satisfy Equation~\ref{eqn:co}.  

There are a variety of values for $v_{eq} \sin i_{\star}$ and $\lambda$ in the literature that could in principle be used as priors. Keeping in mind our aim to include the effects of differential rotation, we opted not to use $v_{eq} \sin i_{\star}$ and $\lambda$ measurements calculated under the assumption of rigid body rotation as priors. As a result, we set no prior for $\lambda$. We also opted not to include priors based on the  \cite{fares10} values for $v_{eq}$ or $\alpha$ since they are based on the assumption that $i_p = i_{\star}$ and that the local stellar photospheric absorption line profiles (Stokes I) can be approximated by a Gaussian function. In the end, we set a uniform prior on $\alpha$ to constrain it to 0-1, where values outside of this region were forbidden (we excluded negative values as no single main-sequence star to date has been detected with anti-solar differential rotation\footnote{We did explore allowing negative values, but only to confirm that an $\alpha = 0$ boundary condition does not impact the results.}); we also constrain $i_p$, $i_{\star}$, and $\lambda$ to 0-90$\degree$, 0-180$\degree$, and -180-180$\degree$, respectively, to avoid degenerate alignments. 

An adaptive principal component analysis was applied to the chains, which required step jumps to take place in an uncorrelated space; this allowed us to better sample the posterior distribution when non-linear correlations were present between parameters \citep[][and references therein]{bourrier15}. The system was analysed with $\sim$20 chains for each $v_{conv}$ formulation, leading to a total of $5 \times 10^6$ accepted steps (for each formulation). Each chain was started at random points near the expected values from the literature. All chains converged to the same solution, and the converged sub-chains were thinned using the correlation length. Finally, we merged the thinned chains, leaving $>10^5$ independent samples of the posterior distribution.

Once the MCMC analysis is complete, the recovered stellar inclination and projected obliquity can be combined with the known orbital inclination to determine the true 3D obliquity ($\psi$) as follows: 
\begin{equation}
\label{eqn:psi}
\psi = \cos  (\ \sin i_{\star}\cos \lambda \sin i_p + \cos i_{\star} \cos i_p \  )^{-1}. 
\end{equation}
Note that Equation~\ref{eqn:psi} can be obtained from the normal vector to the stellar spin-axis in a reference frame created by taking the original XYZ coordinate frame and rotating it about the x-axis by an angle $\pi/2 - i_p$ (as is shown in Inset (b) in Figure~\ref{fig:refframe} -- see \citealt{fabrycky09} for more details).

\subsection{Simulation Data}
We applied this same technique to model stars created using simulated line profiles. For a forward modelling of spectral line profiles, we considered local-box 3D MHD simulations of the stellar near-surface layers with different average magnetic field strength obtained with the \texttt{MURaM} code \citep{MURaM1,MURaM2,Rempel09}. The MHD simulations were created for a representative K dwarf with average magnetic field strengths of 20, 100, and 500\,G, an effective temperature between 4858 and 4901\,K, and a $\log g =4 .609$ \citep{beeck15a}. The solar abundances by \citet{AnGr89} were assumed for the simulations, but the more recent iron abundance of $\log \epsilon_{\mathrm{Fe}}=7.45$ was used \citep[cf.][]{asplund05}. Effective temperature and surface gravity of the simulations are very close to the observationally determined parameters of HD\,189733.

\begin{table}[h!]
\caption[]{Parameters reported from the VALD database for our Fe\,\textsc{i} lines (assuming solar abundances and the stellar parameters in Table~\ref{tab:fix}).}
\centering
\begin{tabular}{c|c|c|c}
    \hline
    \hline
      Line & Depth &  Land\'{e} Factor & Excit. Potential (eV) \\
    \hline  
	610.8 nm & 0.071 & 1.48 & 4.956 \\ 
	616.5 nm & 0.463 & 0.69 &  4.143 \\
	617.3 nm & 0.682 & 2.5 & 2.223 \\
    \hline
  \end{tabular}
\label{tab:lines}
\end{table}

The 3D atmosphere structure provided by the MHD simulations was used to generate synthetic profiles of three Fe\,\textsc{i} lines -- see Table~\ref{tab:lines} for line parameters from the VALD database \citep[][and references therein]{kupka00}; these lines were chosen as they have been well studied in the literature, they vary in line depth (an indicator of formation height), and are present in the HARPS spectrum. This was done for ten different evenly spaced viewing angles, $\mu=\cos\theta$, applying the \texttt{Spinor} code \citep{spinor1,spinor2}. The spatially resolved synthetic spectral line profiles were averaged over six snapshots of the simulation, resulting in a mean profile of an area that is large compared to the granulation scale but small compared to the size of the star. These local average profiles were used as input for a numerical disc integration, applying the same method as described in \citet{beeck13}. In this method, the visible stellar disc is decomposed into stripes along contours of constant local rotational velocity, $v$, and rings along contours of constant $\mu$ (assuming a perfect sphere). The segments of the (projected) areas are identified by a representative rotational velocity $v_i$ and a representative $\mu_j$, and their (projected) areas $w(\mu_i, v_j)=:w_{ij}$ are numerically approximated. The line profile $F_{\mu,v}(\lambda)$, which is given as function of $\mu$ and $v$ is assumed constant within each of these areas and a superposition $\sum_{i,j}w_{ij}F_{\mu_i,v_j}(\lambda)/\sum_{i,j}w_{ij}$ of the line profiles with $w_{ij}$ as weights results in a disc-integrated spectral line profile. This method can handle differential rotation (the rotational input values for the simulation were determined by the MCMC analysis performed on the observed data in Section~\ref{subsubsec:stellrot}) and was modified to also include a transiting planet at arbitrary positions, partially covering the visible stellar disc. 

Accordingly, we injected a transiting planet with parameters matching those in Table~\ref{tab:fix} into the stellar disc integrations. We considered each combination of a single Fe\,\textsc{i} line and magnetic field strength as separate model stars. The transit was sampled from phase $-$0.017 to  $+$0.017, in steps of 0.001. 

\begin{table*}[t!]
\caption[]{MCMC observational results for HD\,189733 and the dervied 3D spin-orbit obliquity}
\begin{center}
\begin{tabular}{c|c|c|c|c|c|c|c||c}
    \hline
    \hline

$v_{eq}$ (km~s$^{-1}$) &  $i{\star}$ ($\degree$) ~\tablefootmark{a}& $\alpha$ & $\lambda$ ($\degree$)  & c$_1$ (km~s$^{-1}$) & c$_2$ (km~s$^{-1}$) & BIC & $\chi^2$ & $\psi$ ($\degree$)\\
    \hline  

3.25$\pm 0.02^{\rm{b}}$ & 90\tablefootmark{b} & 0$^{\rm{b}}$ & -0.45$\pm 0.17$ & -- & -- & 76.0 & 69.0 & --$^{\rm{b}}$ \\

4.45$^{+0.53}_{-0.40}$ & 92.5$^{+12.2}_{-4.1}$ & 0.28--0.85\tablefootmark{c} & -0.42$^{+0.14}_{-0.15}$ & -- & -- & 70.1 & 56.2 & 6.8$^{+12.0}_{-4.1}$ \\

4.50$^{+0.51}_{-0.49}$ & 92.0$^{+11.0}_{-3.8}$ & 0.30--0.86$^{\rm{c}}$ & -0.44$\pm 0.21$ & -0.01$^{+0.08}_{-0.07}$ & -- & 71.4 & 54.0 & 6.3$^{+12.0}_{-3.4}$ \\

4.46$^{+0.54}_{-0.40}$ & 92.3$^{+12.0}_{-4.1}$ & 0.29--0.85$^{\rm{c}}$ & -1.7$^{+0.89}_{-0.86}$ & 1.0$^{+0.71}_{-0.71}$ & -0.95$^{+0.67}_{-0.62}$ & 74.5 & 53.7 & 6.9$^{+11.0}_{-3.7}$ \\

    \hline
  \end{tabular}
\end{center}
\tablefoot{ \tablefoottext{a}{$i_{\star}$ is constrained to 0-180$\degree$, and values $>90$$\degree$ indicate the star’s rotation axis is pointig away from the LOS.} \tablefoottext{b}{Fixed under the assumption of rigid body rotation; note this means the value in the $v_{eq}$ column for this row corresponds to $v_{eq} \sin i_{\star}$ and that we are unable to determine the 3D obliquity, $\psi$.} \tablefoottext{c}{We present only the 1$\sigma$ ranges as $\alpha$ is largely unconstrained by the data; however, we did find $\alpha$ is $>0.1$ with 99.2\% confidence.}}
\label{tab:mcmc}
\end{table*}

Note, since we simulated only three Fe\,\textsc{i} lines, it is entirely possible that our observed CCFs may not capture and preserve the behaviour of these individual lines. If this is the case, then a comparison between simulation and observation is futile. Consequently, we examined CCFs created from a variety of template masks, in addition to the standard HARPS pipeline mask mentioned in Section~\ref{sub:obs}. Unfortunately, the reduction in the number of lines in these template masks significantly decreased the overall RV precision. This led to increased scatter in the observed data that prevented conclusive analysis; as such, we do not discuss alternate template masks in the remainder of the paper. 

\section{Empirical Rotation and Obliquity Results}
\label{sec:results}
\subsection{MCMC Posterior Probability Distributions}
\label{subsec:mcmc}
We ran four sets of MCMC chains to model the $v_{tot}$, corresponding to the three polynomial configurations for the $v_{conv}$ component (0th-2nd order) and one rigid body configuration for $v_{stel}$ (note because we subtracted off the out-of-transit master CCF RVs, the zeroth order scenario may represent either no convective contribution or a convective contribution that does not vary across the stellar disc). The model with rigid body stellar rotation followed the same $v_{conv}$ formulation as the best-fit model with differential rotation.

These four sets were composed of $\sim$20 MCMC chains each, with $\sim 10^5$ accepted steps and an acceptance rate of $\sim$ 20-30\%. The best fit values for the $v_{tot}$ model parameters were inferred from the medians of the posterior probability distributions and are given in Table~\ref{tab:mcmc}, alongside the derived 3D spin-orbit obliquity ($\psi$) and 1$\sigma$ errors that were evaluated by taking limits at 34.1\% on either side of the median. Note that because the posterior probability distributions for $\alpha$ were broad and relatively flat, we present only the 1$\sigma$ ranges and refrain from quoting a best-fit value. The posterior probability distributions for the $v_{tot}$ model parameters, when considering a zeroth order polynomial $v_{conv}$ are shown in Figure~\ref{fig:mcmc_dist_noCB}, along with the marginalised 1D distributions. The posterior probability distributions for the higher order $v_{conv}$ formulations can be found in Appendix~\ref{apn:mcmc}. To distinguish between the different $v_{conv}$ models we calculated the $\chi^2$ and the Bayesian information criterion (BIC), shown in Table~\ref{tab:mcmc}. We found the improvement in the $\chi^2$ for the linear and quadratic $v_{conv}$ was not enough to offset the increase in the BIC, and therefore conclude the best fit for this data is the constant $v_{conv}$ formulation. The $\chi^2$ and the BIC also both indicated that the model with differential stellar rotation was a better fit to the data than the model with rigid body rotation. 

Regardless of the $v_{conv}$ formulation (best fits shown in Figure~\ref{fig:mcmc_bestfit} and discussed in Section~\ref{subsec:CB}.), the recovered stellar rotation parameters ($v_{eq},\ i_{\star},\ \alpha,\ \lambda$, and derived $\psi$) were consistent with one another (within 1-2$\sigma$), with strong correlations between $v_{eq}$ and $\alpha$ (there were also correlations between $\psi$ and $i_{\star}$, but this is because $\psi$ depends directly on $i_{\star}$ -- see Equation~\ref{eqn:psi}). However, correlations between the equatorial velocity and the differential rotation rate are expected for systems that are closely aligned because the planet does not transit enough stellar latitudes to independently determine these quantities. 

\begin{center}
\begin{figure*}[t!]
\centering
\includegraphics[trim=1.9cm 12.85cm 6.9cm 2.5cm, clip, scale=1.3]{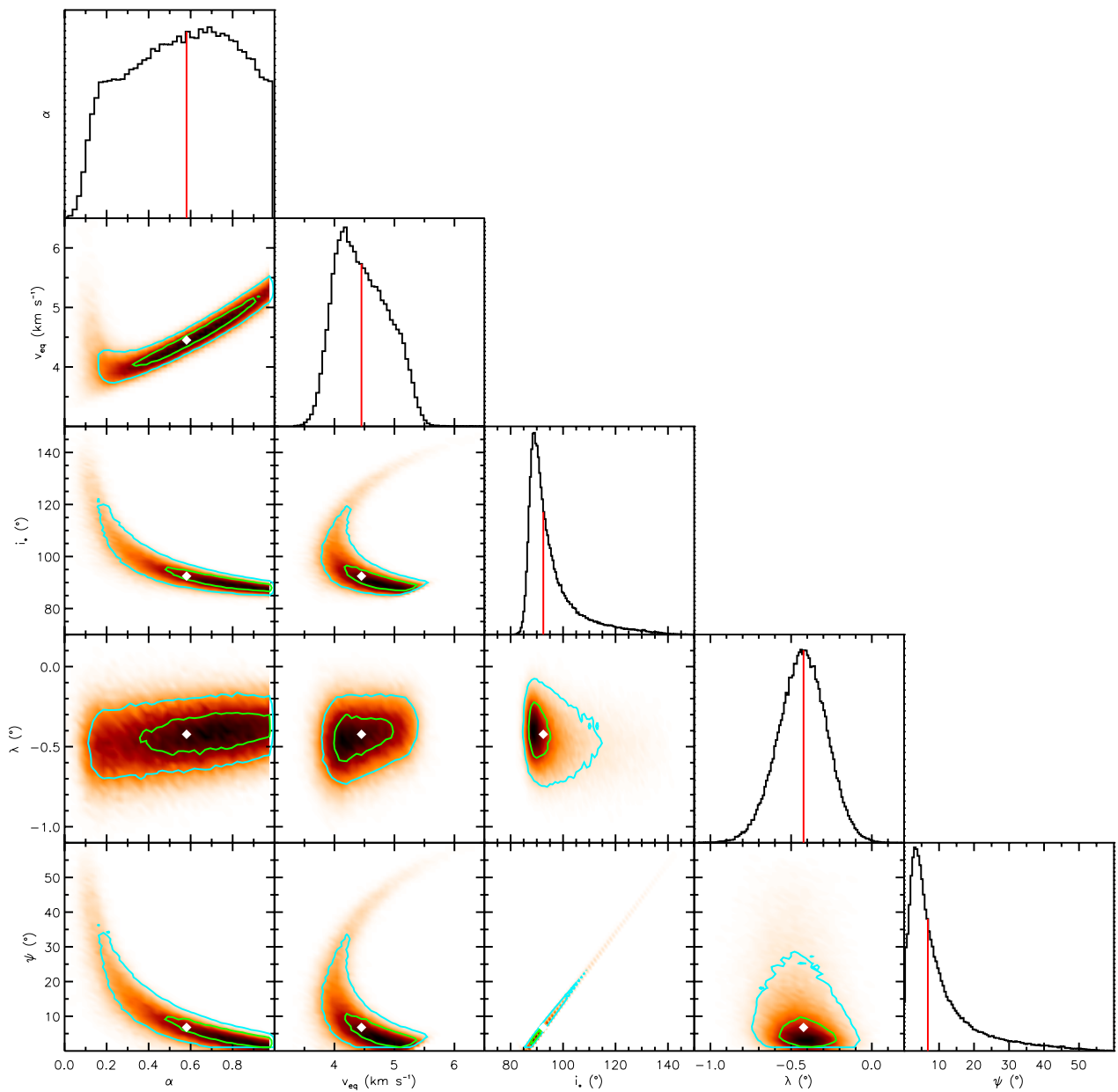}
\caption[]{Correlation diagrams for the probability distributions of the $v_{tot}$ model parameters, when the $v_{conv}$ contribution is set to 0.  Green and blue lines show the 1 and 2$\sigma$ simultaneous 2D confidence regions that contain respectively 39.3\% and 86.5\% of the accepted steps. 1D histograms correspond to the distributions projected on the space of each line parameter. The red line and white point show median values. Note $\psi$ is derived from Equation~\ref{eqn:psi} and is not an MCMC jump parameter.} 
\label{fig:mcmc_dist_noCB}
\end{figure*}
\end{center}
\vspace{-30pt}

\subsection{Comparisons to the Literature}
\label{subsec:lit}
Due to the $\alpha - v_{eq}$ degeneracies in this dataset, it is more appropriate to compare the product $v_{eq} \sin i_{\star} (1 -  \alpha~y_{\perp}'^2)$ to the previous $v_{eq} \sin i_{\star}$ quoted for HD\,189733 in the literature (rather than medians of the $v_{eq}$ or $\alpha$ distributions). For example, if we substitute $y_{\perp}'$ with the known impact factor, b, (which will be reasonably close to $y_{\perp}'$ since the true  obliquity is close to zero) we find $v_{eq} \sin i_{\star} (1 -  \alpha~b^2) \approx 3.3$~km~s$^{-1}$; remarkably, this value is equal to the $v_{eq} \sin i_{\star}$ recovered by \cite{triaud09} before they adjusted their RM model to try to account for underlying trends in their residuals. This is also consistent with the $v_{eq} \sin i_{\star}$ reported in the literature for line broadening and spectropolarimetric techniques (see Table~\ref{tab:vsini}), but is greater than the values reported from line profile tomography and model stars that have been adjusted to reduce the residuals. Hence, the assumption of rigid body stellar rotation may be biasing $v_{eq} \sin i_{\star}$ towards lower values (note this is indeed found in our results under the rigid body assumption, though to a lesser extent than in the literature -- perhaps because we analyse the local residual CCF RVs directly). 

We remind the reader that due to the degeneracies in this system, the median of the $\alpha$ posterior probability distribution is unlikely to give the true differential rotation rate for this star. Indeed, the posterior distributions demonstrate that $\alpha$ is relatively unconstrained by the data, but we can effectively rule out rigid body rotation as $\alpha$ is $>0.1$ with 99.2\% confidence (and $>$ 0.2 with 91.7\% confidence). This is in agreement with the results from \cite{fares10} that indicate $\alpha \approx 0.278 \pm 0.093$. 

\begin{table}[t!]
\caption[]{Previous $v_{eq} \sin i_{\star}$ reported for HD\,189733}
\centering
\begin{tabular}{c|c|c}
    \hline
    \hline
$v_{eq} \sin i_{\star}$ (km~s$^{-1}$) & Method & Reference\\
    \hline  

3.5$\pm 1$ & Line Broadening & \citealt{bouchy05} \\
2.97$ \pm 0.22$ & Model Star & \citealt{winn06} \\
3.2$\pm 0.7$ & Line Broadening & \citealt{winn06}\tablefootmark{a} \\
3.316$^{+0.017}_{-0.067}$ & Model Star & \citealt{triaud09} \\
3.05 & Model Star\tablefootmark{b} & \citealt{triaud09} \\
3.08 - 3.1$\pm 0.02$\tablefootmark{c} & Tomography & \citealt{cameron10} \\
3.41$\pm 0.02$\tablefootmark{d}& Polarimetry & \citealt{fares10} \\

    \hline  
  \end{tabular}
\tablefoot{\tablefoottext{a}{D. Fischer 2006.} \tablefoottext{b}{Adjusted to reduce trend in residuals.} \tablefoottext{c}{Exact result was limb darkening and transit dependent.} \tablefoottext{d}{Converted from their reported equotorial period, under their assumption $i_{\star} \approx i_p$.}}
\label{tab:vsini}
\end{table}

Moreover, the sky-projected obliquities are also inline with previously reported values in the literature \citep[$-1.4 \pm 1.1\degree$ to $-0.35 \pm 0.25\degree$;][]{winn06,triaud09,cameron10}. A small misalignment in the sky-projected obliquities indicates that the stellar inclination is statistically most likely to be near the orbital inclination. This is indeed what we found ($i_{\star} \approx 92^{+12}_{-4}$$\degree$, indicating the star is pointing slightly away from the LOS), with the stellar inclination constrained with good precision. The combination of both the sky-projected obliquity and the stellar inclination, allowed us to self-consistently recover, for the first time, the \textit{true 3D spin-orbit geometry} of the HD\,189733 system. We found $\psi \approx 7^{+12}_{-4}$$\degree$, which indicates that the orbit of HD\,189733\,b is indeed (mostly) aligned with its host star's stellar spin-axis. Note, this is in agreement with the 3D obliquity of $\psi = 4^{+18}_{-4}$$\degree$ obtained by \cite{dumusque14}, where they analysed starspot signatures in conjunction with the projected obliquity from the \cite{triaud09} RM modelling.

From the $v_{tot}$ model fits in Figure~\ref{fig:mcmc_solid_diff}, we also found that with this new technique and accounting for differential rotation, we did not see the wave-like pattern displayed in the residuals as previous authors have reported \citep[e.g.][]{triaud09}. However, when we assumed rigid body stellar rotation, we did see a hint of a wave-like pattern in our residuals; therefore such a pattern in the residuals may indicate non-negligible differential stellar rotation is contributing to the observed RVs.  

\subsection{Robustness of the Best fit Models}
\label{subsec:robust}
We also analysed each night independently for our best fit model (i.e. with differential rotation and constant $v_{conv}$), to ensure that the results were not dominated from a single night. We found that all the fitted parameters agreed well within 1$\sigma$, and that each night strongly favoured $\alpha > 0$. However, we note that the $\alpha$ distribution, while still broad, did have a stronger peak closer to 0.2 in the Sept. 7 and Aug. 28 data as compared to the July 29 data. This difference could potentially be linked to the level of magnetic activity; for a stronger magnetic field, we would expect less convective induced redshift near the limb (see Section~\ref{subsec:CB}) and therefore more blueshifted RVs at ingress.  

Additionally, we also tested the impact of imposing stricter $\mu$ cuts (note we did not relax the cuts to values $<0.25$ due to the SNR therein). We found that because stricter limb cuts limit the analysis to fewer stellar latitudes (see inset in Figure~\ref{fig:mcmc_solid_diff}) it pushed the solutions closer to rigid body rotation. For example, we found with a $\mu$ cut of 0.5 we could no longer distinguish between rigid body and differential stellar rotation. Hence we argue, at least for nearly aligned systems, that the $\mu$ constraints should be as relaxed as possible.

As mentioned in Section~\ref{subsec:tech}, we also explored the impact of limb darkening by testing extremes based off the R and B bands and found all fit parameters agreed with our best fit model within 1-2$\sigma$. However, we note here that there was a slight trend for larger RVs near the limb when stronger limb darkening was considered; such a trend could potentially affect observations with more precise future instruments and/or stars with different stellar rotation properties.  

\begin{center}
\begin{figure}[t!]
\centering
\includegraphics[trim=0.5cm 0.25cm 0.25cm 0.4cm, scale = 0.45]{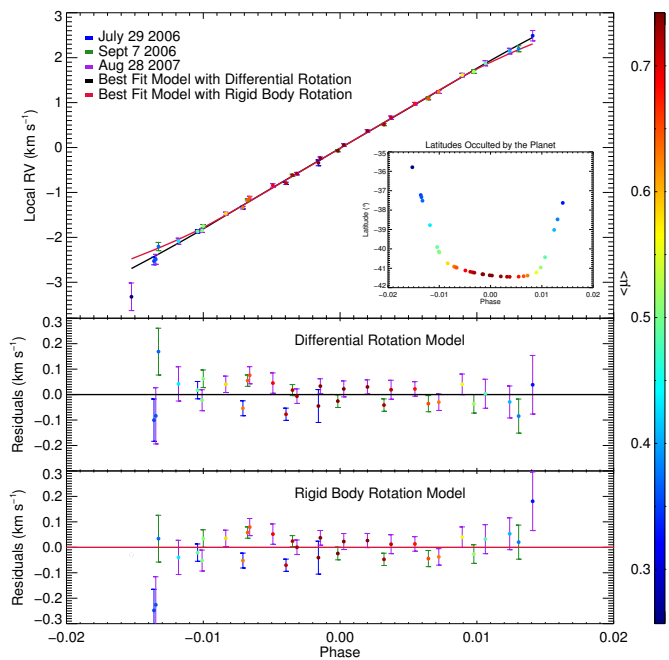}
\caption[]{Top: Net velocity shifts of the in-transit residual CCFs as a function of phase, alonside the best fit models (i.e. constant $v_{conv}$) for rigid body (red) and differential (black) rotation. Middle and Bottom: Residuals (local RV - model $v_{stel}$) for differential and rigid body stellar rotation, respectively (horizontal lines at 0 to guide the eye); note the point at phase $=$ -0.15 with the large error is not displayed in order to better view the remaining points. The colour-coding of the data points indicates the stellar disc position (brightness-weighted $\left < \mu \right >$ behind the planet), while the colour of the error bar indicates the observation date. Inset: Occulted stellar latitudes determined by the best fit differential rotation model.} 
\label{fig:mcmc_solid_diff}
\end{figure}
\end{center}
\vspace{-40pt}

\begin{center}
\begin{figure}[t!]
\centering
\includegraphics[trim=0.5cm 0.25cm 0.25cm 0.6cm, scale = 0.45]{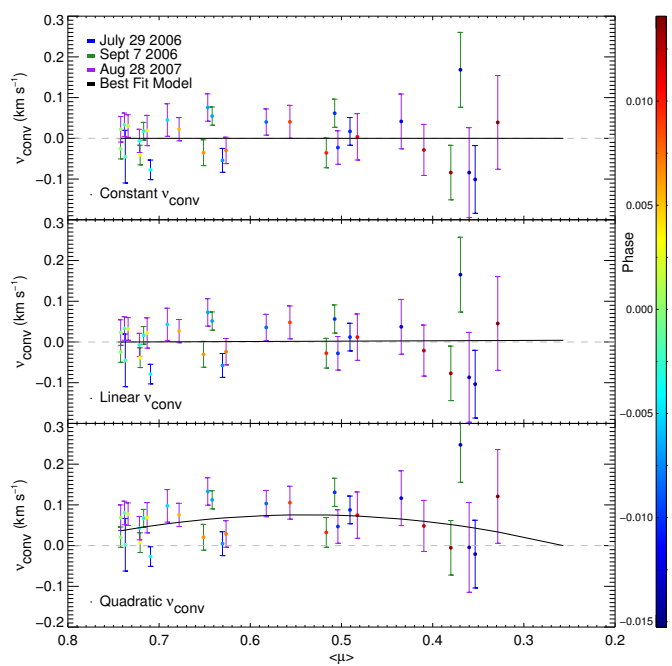}
\caption[]{Net convective velocity shifts determined from subtracting the $v_{stel}$ model fits from the local RVs of the in-transit residual CCFs as a function of stellar disc position, defined as the brightness-weighted $\left < \mu \right >$ behind the planet. Shown for each $v_{conv}$ formulation: constant (top), linear (middle), and quadratic (bottom); note a point at $\left < \mu \right > \approx$ 0.26 (and $v_{conv} \approx$ -0.54 -- -0.62 km~s$^{-1}$) with large error is not shown in order to better view the remaining points. The colour-coding of the data points indicates the phase, while the colour of the error bar indicates the observation date. Horizontal dashed lines at 0 are shown to guide the eye.} 
\label{fig:mcmc_bestfit}
\end{figure}
\end{center}

\section{Empirical and Simulated Local Photospheric Profile Variation Results}

\subsection{Convective blueshift Across the Stellar Limb}
\label{subsec:CB}
We are also interested in how the net convective blueshift (CB) varies across the stellar limb. This is because solar observations (and simulations) indicate the net CB decreases from disc centre to limb \citep[on the 100s of m~s$^{-1}$ level;][]{dravins82}, and can even result in a net redshift. As stated in Section~\ref{sec:intro}, such a variation is expected due to geometrical effects; towards the stellar limb different aspects of the granulation fall along our LOS (e.g. near the limb, the granular walls become visible and the granule peaks and bottoms of the lanes are hidden). The result is different brightness ratios and LOS RVs (e.g. flows orthogonal to the granular peaks have a LOS component). The decrease in CB towards the limb arises because the redshifted flows are more often observed in front of the hotter plasma above the intergranular lanes \citep[see][and references therein, for more details]{balthasar85, asplund00}. This centre-to-limb CB variation is also expected in K dwarfs. Since the granule to intergranular lane contrast is less and the flow velocities are lower in K dwarfs, this variation may be lower in HD\,189733 than the Sun.

To measure the net convective velocity shift of the residual CCFs (and simulated line profiles), we first removed the stellar rotation contribution from the measured RVs. This was done by calculating $v_{stel}$ in Equation~\ref{eqn:vstel} using the $v_{eq}, \ i_{\star}$, $\alpha$, and $\lambda$ obtained through the MCMC analysis in Section~\ref{sec:results}. In Figure~\ref{fig:mcmc_bestfit} we show these results for each $v_{conv}$ formulation, alongside the best-fit $v_{conv}$ polynomials (one point at  $\left < \mu \right > \approx$ 0.26, and $v_{conv} \approx$ -0.54 -- -0.62 km~s$^{-1}$ with a large error is excluded from view). Note, the observed RVs are relative to the net CB of the out-of-transit CCFs, and therefore we cannot comment on the absolute blue- or redshift. 

As shown in Table~\ref{tab:mcmc}, with increasing polynomial order, the $\chi^2$ decreased, but the BIC increased. Hence, the improvement in the fit for the higher order $v_{conv}$ is not sufficient enough to justify the extra free parameters. Hence, the best-fit is a constant offset and any CB variation is not significantly larger than the error on the residual CCF RVs ($\sim 50$~m~s$^{-1}$). Nonetheless, we note that both the linear and quadratic $v_{conv}$ fits predict a slight decrease in CB away from mid-transit positions; however, the quadratic fit also indicates an increase in CB near the limb, primarily due a single outlier with high blueshift (and high error) at $\left < \mu \right > \approx 0.26$.

We also compared the observed net convective shifts (for the constant $v_{conv}$) to the MHD simulations in Figure~\ref{fig:CB}. We remind the reader that the residual CCFs are relative to the master out-of-transit CCFs; to put the simulations on the same scale, we subtracted the RVs from the out-of-transit model stars. Since, the planet covers a large range of $\mu$ during a given exposure there is not a one-to-one relationship between observation and simulation. To illustrate this, we plotted three representative error bars based on the range of brightness weighted $\mu$ covered during a 450~s exposure. This includes a very strong assumption that the residual profiles vary close to linearly in $\mu$ during an exposure; which may not be the case, but serves as our best approximation.

\begin{center}
\begin{figure}[t!]
\centering
\includegraphics[trim=0.cm 0.25cm 0.25cm 0.5cm, clip, scale=0.445]{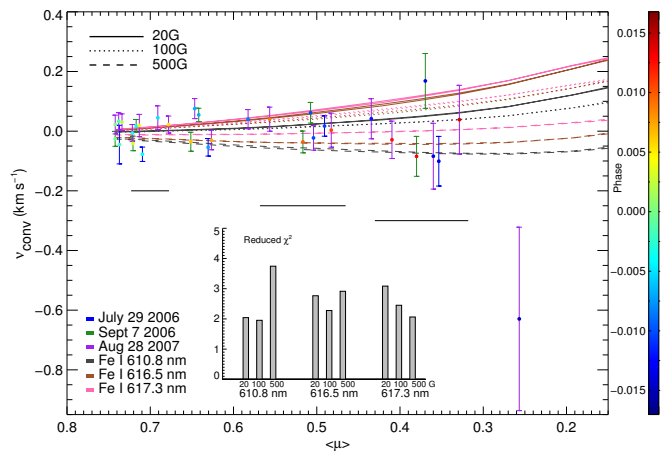}
\caption[]{Net convective velocities of the residual CCF/line profiles. The Fe\,\textsc{i} 610.8 nm, 616.5 nm, and 617.3 nm simulated data are plotted as grey, brown, and pink lines, respectively; line style indicates the average magnetic field strength, with solid, dotted, and dashed lines representing 20, 100, and 500~G simulations, respectively. HD\,189733 data are colour-coded by phase, with error bars colour-coded by observation night. Black horizontal bars represent the estimated error on the observed $\left < \mu \right >$ at different locations. Inset displays the reduced $\chi^2$.} 
\label{fig:CB}
\end{figure}
\end{center}
\vspace{-30pt}

\begin{center}
\begin{figure}[t!]
\centering
\includegraphics[trim=0.5cm 0.25cm 0.25cm 0.5cm, clip, scale=0.445]{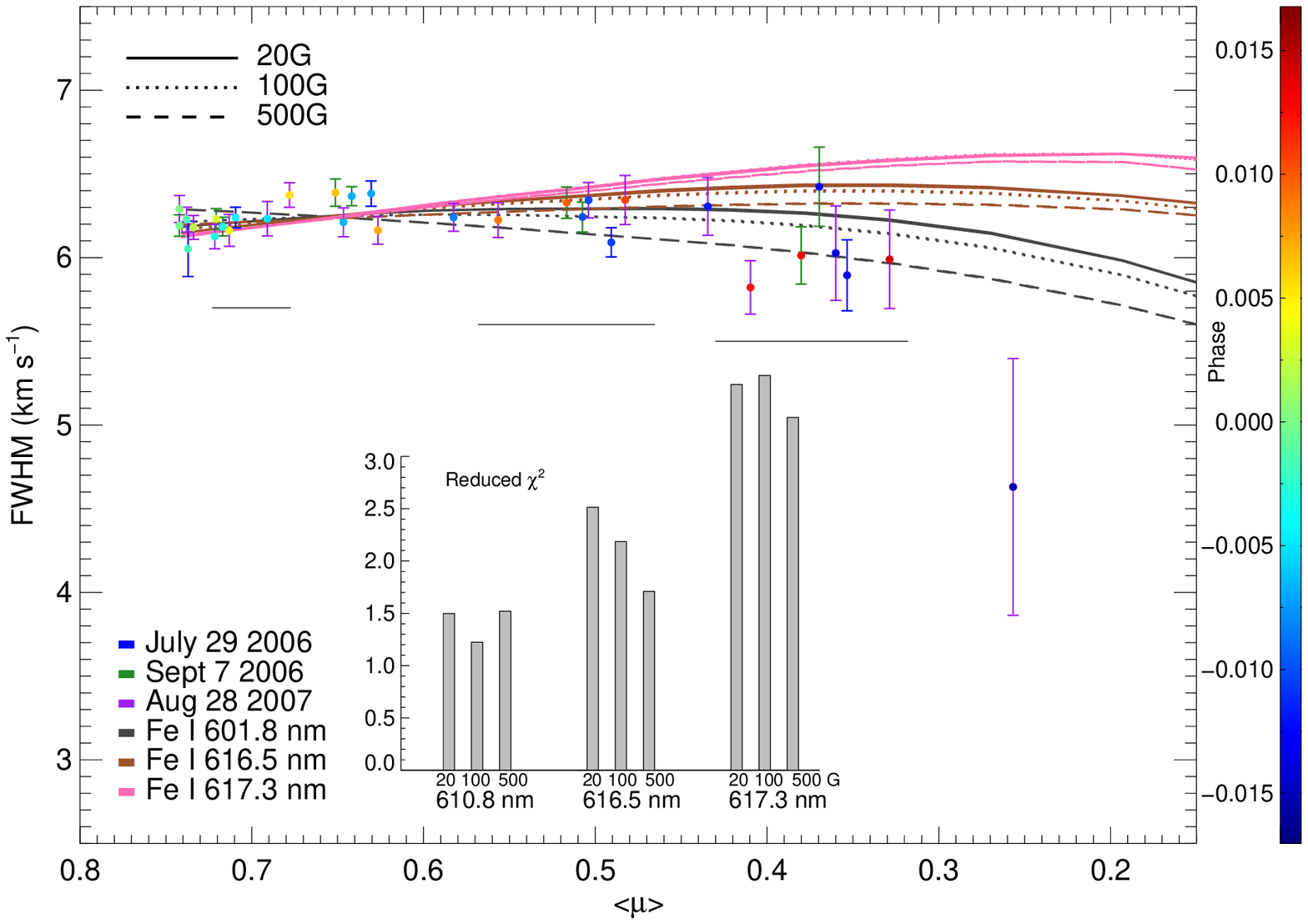}
\includegraphics[trim=0.4cm 0.25cm 0.25cm 0.5cm, clip, scale=0.445]{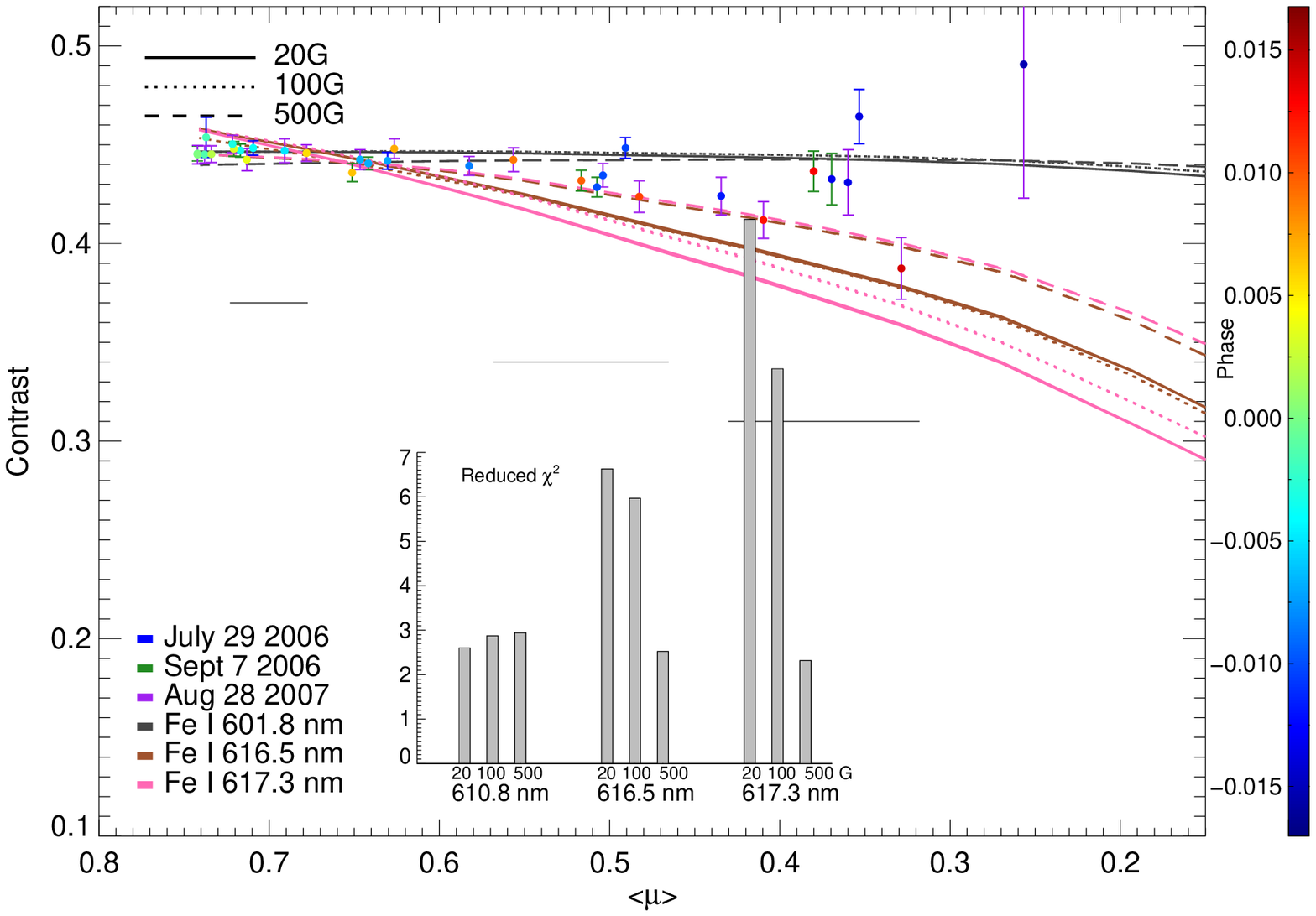}
\includegraphics[trim=0.4cm 0.25cm 0.25cm 0.5cm, clip, scale=0.445]{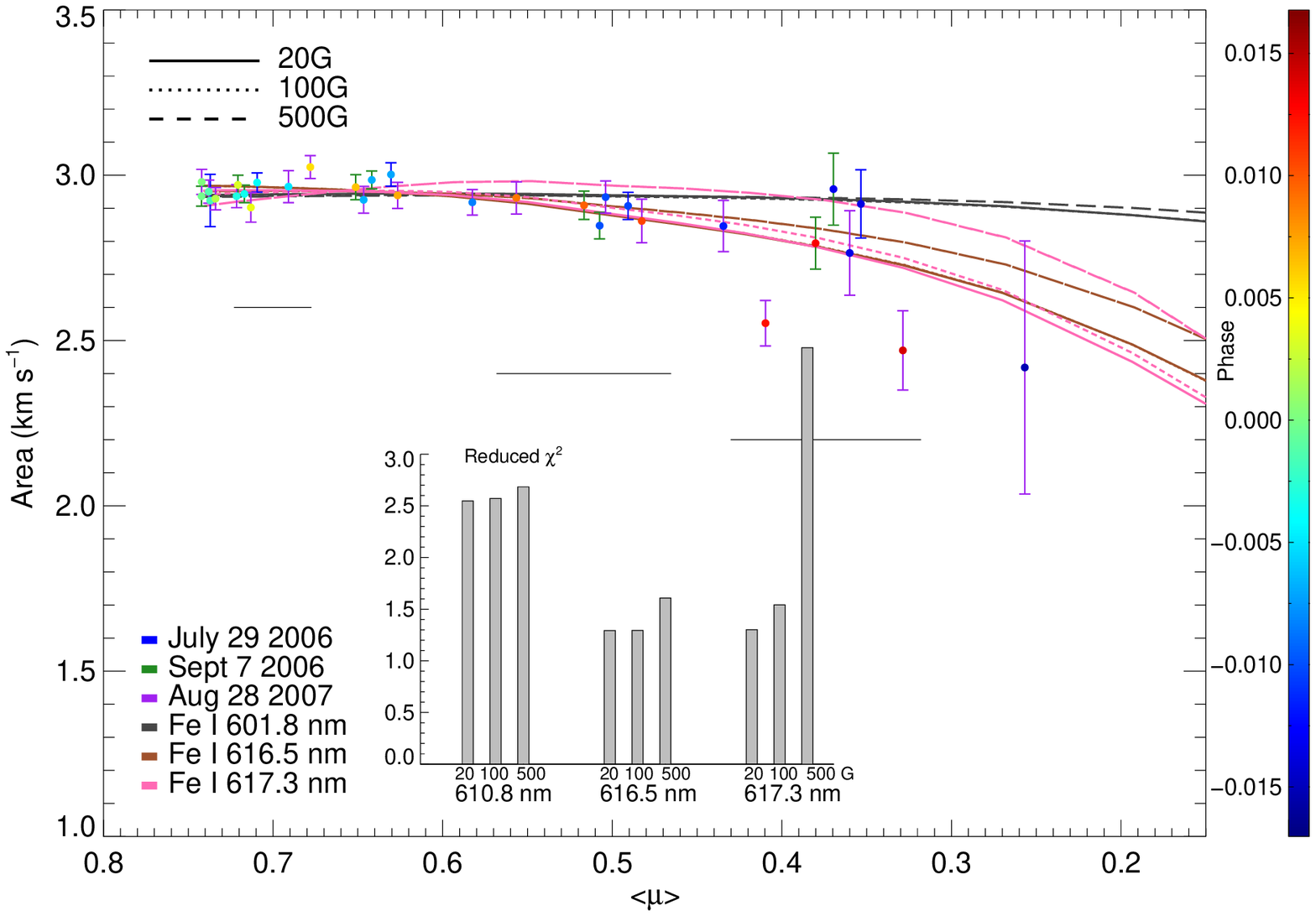}
\caption[]{FWHM (Top), contrast (Middle), and area (Bottom) of a Gaussian fit to the residual profiles. The Fe\,\textsc{i}  610.8 nm, 616.5 nm, and 617.3 nm simulated data are plotted as grey, brown, and pink lines, respectively; line style indicates the average magnetic field strength, with solid, dotted, and dashed lines representing 20, 100, and 500~G simulations, respectively. HD\,189733 data are colour-coded by phase, with error bars colour-coded by observation night. Black horizontal bars represent the estimated error on the observed $\left < \mu \right >$ at different locations. Inset displays the reduced $\chi^2$.} 
\label{fig:fwhm_con_area}
\end{figure}
\end{center}
\vspace{-30pt}

The data exhibited no CB variation, but was consistent with the radiative 3D MHD simulations. It is important to note that since HD\,189733 is an active star, the planet may have transited regions of different magnetic field strength, which could be responsible for some of the scatter seen in the observed data. Due to this scatter, the best reduced $\chi^2$ is $\sim 2$. Overall, the simulated data predicts (for a K dwarf) a redshift near the limb that increases with increasing line depth and decreasing magnetic field. Note that the simulations with the highest magnetic field either experience no variation or a slight blueshift until $\left < \mu \right > \approx 0.25$, before they start to redshift.  These results indicate that convection effects may be negligible for magnetically active K dwarfs with HARPS-level precision. However, these effects grow with decreasing magnetic field, and (from our knowledge of the Sun) we also expect these effects to be larger in G dwarfs. Hence, such effects may not be negligible with future instrumentation and/or for hotter, less active stars.

\subsection{Local CCF Shape Across the Stellar Limb}
We also analysed the centre-to-limb shape changes of the local CCFs/line profiles by examining the FWHM, contrast, and area of a Gaussian fit to the residual profiles (this approach was taken to match the analysis of typical exoplanet observations); these are shown in Figure~\ref{fig:fwhm_con_area}. The comparison between observed and simulated data serves to check the validity of the 3D MHD simulations and provide insight into the origin of any observed variations. We remind the reader that the simulated data originate from a single line profile, while the observations consist of a CCF. Accordingly, the absolute values of the shape diagnostics will not be the same for the simulated and observed residual profiles. Instead, it is the variation in the shape across the stellar limb that should be consistent (if the CCF preserves the behaviour of these Fe\,\textsc{i} lines). For this reason, the shape diagnostics from the simulated data were shifted to minimise the $\chi^2$ with the observations. Since the simulated transits were sampled more finely (and evenly) in $\left < \mu \right >$, we fit a second order polynomial to them to calculate the $\chi^2$. 

For the simulated star, we include stellar rotation of HD\,189733 from the best-fit MCMC analysis. However, because the region behind the planet is small, the impact of stellar rotation is minimal; hence, even if our (degenerate) recovered $\alpha$ is high, this does not significantly alter the conclusions about the centre-to-limb shape variations as they are primarily dominated by the convective properties. The observational data is independent of our modelling of the stellar rotation. 

There are many reasons to expect a centre-to-limb shape variation in the local profile/CCF. For example, towards the limb some convective flows will be be hidden behind granules, which result in a reduction in the line broadening induced by the LOS velocities and could decrease the FWHM \citep{beeck15b}. However, one could also expect an increase in the FWHM because of an increasing contribution of the flows orthogonal to the granule peaks, since these flows have a higher variability than those parallel \citep{beeck13}. 

Additionally, the area and contrast could decrease towards the stellar limb as forefront granules in the forefront obscure the background granules and intergranular lanes. Naively, one might expect lines with a deeper formation height to have a larger decrease in area and contrast across the stellar limb as they may be more affected by the granulation geometry; however, this is a very simplistic view as the line behaviour depends on more than the formation height (e.g. magnetic and temperature sensitivity, ionisation/excitation potential etc).

Interestingly, the observed data exhibits little shape variation across the stellar disc; however, given the error, this is roughly consistent the MHD-based model stars. For example, both the FWHM and area diagnostics achieve reduced $\chi^2 \approx 1$ between the observed data and at least one simulated line. The comparison with contrast is more discrepant, but still achieves a reduced $\chi^2 \approx 2$ for many cases. Moreover, this discrepancy may arise because the construction of the CCF does not preserve the effective line depth of the stellar spectrum and therefore may not preserve the expected variation in contrast across the stellar limb. 

Overall, the simulated lines showed a decrease in FWHM gradient for both shallower lines and higher magnetic field strength. The simulated lines also showed trends in contrast and area; wherein an increase in line depth and magnetic field led to a decrease in the absolute value of gradient, which eventually reached $\sim 0$ for the weakest line. Consequently, the lack of variations seen in the observed data are likely because HD\,189733 is a magnetically active K dwarf. Furthermore, larger centre-to-limb variations are expected for both less active stars and for hotter spectral types.

\section{Concluding Remarks}
\label{sec:conc}
To date, planetary migration still remains an open question within the exoplanet community. To solve this, exoplanetary migration theories must be empirically validated. One of the best ways to distinguish between such theories is by determining the alignment between a planet's orbital plane and the host star's rotation axis. This can be done by modelling the observed RM effect during a planetary transit. However, we must be careful to include all the relevant physics in the RM modelling in order to avoid introducing signifiant biases into our analysis. Additionally, we must disentangle the true 3D obliquity from the sky-projected obliquity traditionally measured through the RM effect, lest we introduce additional biases. 

Throughout this paper, we have presented a new RM modelling technique that directly measures the local CCF and RV of the stellar surface occulted during a planetary transit. Moreover, it circumvents many assumptions on the behaviour of the stellar photosphere and is capable of directly determining the 3D obliquity for many star-planet systems. The highlights of this new technique are summarised below: 

\begin{itemize}
\item Since the local photospheric CCF is directly measured no assumptions are made on the shape of local profile or the disc-integrated profile. 

\item Convective contributions from the stellar photosphere can be accounted for, including centre-to-limb variations in the local profile shape and net convective blueshift.  

\item If the planetary orbit is even slightly misaligned with the rotation axis, then we can directly probe differential stellar rotation. 

\item Hence, for numerous systems we can model the differential rotation, solve for the stellar inclination, lift the $v_{eq} \sin i_{\star}$ degeneracy, and determine the true 3D obliquity (without the need for complementary techniques). 

\item It is efficient and requires few free parameters as many terms can be fixed from a high precision transit light curve. 
\end{itemize}

We note that this technique is currently limited to systems with high precision spectrographic observations (both in terms of high spectral resolution and precise RVs). It also requires a number of parameters from a transit light curve ($a/R_{\star}$, $i_p$, $R_p/R_{\star}$, and limb darkening coefficients) and the planetary orbital motion from modelling sufficient out-of-transit RVs (though these could be added as extra free parameters in the model); however, any systems that would be suitable for this analysis will likely have these readily available.

Herein, we have applied this new technique to the transit of HD\,189733\,b and compared the results to 3D MHD simulations. We summarise these findings below: 

\begin{itemize}
\item Rigid body rotation can be excluded at high confidence ($>99$\% probability that $\alpha > 0.1$). 

\item The rigid body rotation assumption biases the projected equatorial velocity toward lower values. 

\item Modelling the stellar rotation as rigid-body may be the culprit behind the wave-like residuals seen in previous RM modelling of this system. 

\item The stellar rotation modelling is largely independent of convection for this system. However, we note that the $v_{conv}$ polynomial coefficients were all highly correlated with one another and with the projected obliquity, which suggests a potential degeneracy between the these parameters. 

\item We recovered a sky-projected obliquity of $\lambda\approx-0.4 \pm 0.2$$\degree$ and a 3D obliquity of $\psi \approx 7^{+12}_{-4}$$\degree$. 

\item This is the first time the 3D obliquity has been self-consistently measured for HD\,189733\,b, and only the fourteenth 3D obliquity measured in exoplanet systems.\footnote{TEPCAT catalogue (http://www.astro.keele.ac.uk/jkt/tepcat/rossiter.html).}

\item The observed local CCFs exhibit no significant shape change or convective blueshift variations across the stellar disc. Within the error, this is in agreement with predictions from 3D MHD simulations for a magnetically active K dwarf. 

\item The 3D MHD simulations predict various trends based on the average magnetic field strength and the depths of the line profiles. Such as, a decrease in the centre-to-limb variation of the convective blueshift with increasing magnetic field and decreasing line depth.

\end{itemize}

In forthcoming publications we aim to apply this new technique to a variety of (bright) stars with transiting planets. In particular, we will target systems with higher predicted star-planet misalignments as misaligned systems will allow us to better constrain the differential rotation. Preliminary simulations also indicate a decrease in the degeneracy of the recovered $v_{conv}$ polynomial and projected obliquity. We will also target stars of varying spectral type and magnetic activity, as current 3D MHD simulations predict stellar surface magnetoconvection will have an larger impact on the RM modelling for both hotter stars and less active stars. 

Throughout this paper we have shown that with HARPS-level precision, we have the ability to directly measure the stellar photosphere at multiple centre-to-limb positions and that this information can be used to more accurately model the RM effect. Such improved accuracy will only become more important as future instruments promise even higher precision (i.e. ESPRESSO on the VLT and EPDS on the WIYN telescope), and will therefore require a more detailed modelling of the stellar surface than previous instruments.

\begin{acknowledgements} 
We thank the referee for their considered report, which led to important conclusions on the robustness of the results herein. HMC would like to thank R. D. Brothwell, D. J. A. Brown, E. de Mooij, and S. Moutari for useful discussions. This work bas been carried out in the framework of the National Centre for Competence in Research ``PlanetS'' supported by the Swiss National Science Foundation (SNSF). HMC, CL, VB, and FP acknowledge the financial support of the SNSF. HMC and CAW gratefully acknowledge support from the Leverhulme Trust (grant RPG-249). CAW also acknowledges support from STFC grant ST/L000709/1. BB acknowledges research funding by the Deutsche Forschungsgemeinschaft (DFG) under the grant SFB~963/1, project A16. This research has made use of NASA's Astrophysics Data System Bibliographic Services, and the VALD database, operated at Uppsala University, the Institute of Astronomy RAS in Moscow, and the University of Vienna.
 \end{acknowledgements}

\bibliographystyle{./bibtex/aa}
\bibliography{abbrev,mybib}

\begin{appendix}

\section{Additional Posterior Probability Distributions}
\label{apn:mcmc}

\begin{center}
\begin{figure*}[b!]
\centering
\includegraphics[trim=1.9cm 12.75cm 6.75cm 2.5cm, clip, scale=1.4]{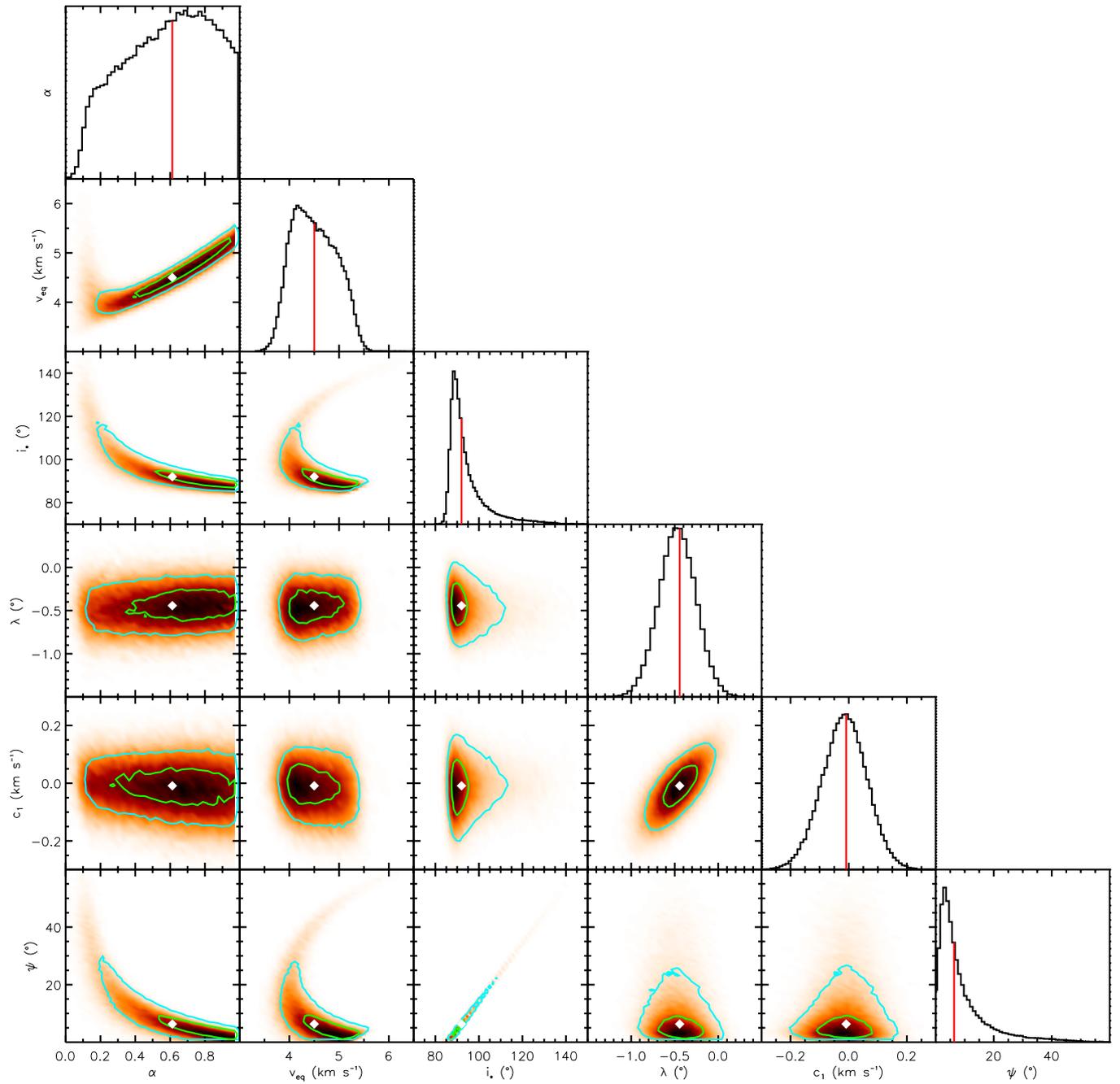}
\caption[]{Correlation diagrams for the probability distributions of the $v_{tot}$ model parameters, for the first order $v_{conv}$ function. Green and blue lines show the 1 and 2$\sigma$ simultaneous 2D confidence regions that contain respectively 39.3\% and 86.5\% of the accepted steps. 1D histograms correspond to the distributions projected on the space of each line parameter. The red line and white point show median values. Note $\psi$ is derived from Equation~\ref{eqn:psi} and is not an MCMC jump parameter.} 
\label{fig:mcmc_dist_linCB}
\end{figure*}
\end{center}

\begin{center}
\begin{figure*}[b!]
\centering
\includegraphics[trim=1.9cm 12.75cm 6.75cm 2.5cm, clip, scale=1.4]{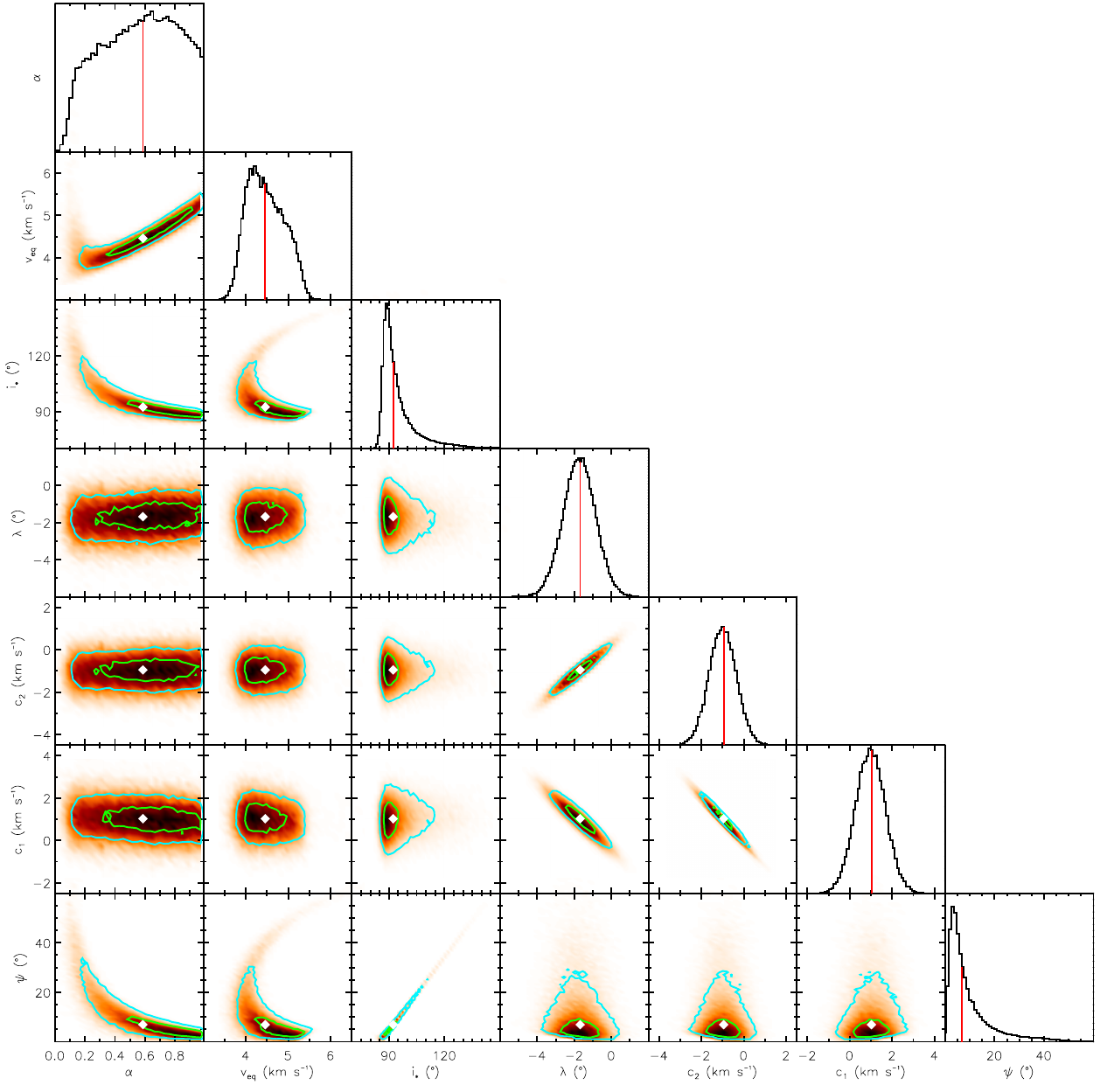}
\caption[]{Correlation diagrams for the probability distributions of the $v_{tot}$ model parameters, for the second order $v_{conv}$ polynomial. Green and blue lines show the 1 and 2$\sigma$ simultaneous 2D confidence regions that contain respectively 39.3\% and 86.5\% of the accepted steps. 1D histograms correspond to the distributions projected on the space of each line parameter. The red line and white point show median values. Note $\psi$ is derived from Equation~\ref{eqn:psi} and is not an MCMC jump parameter.} 
\label{fig:mcmc_dist_2CB}
\end{figure*}
\end{center}

\end{appendix}

\end{document}